\documentclass[acmlarge]{acmart}
\AtBeginDocument{%
  \providecommand\BibTeX{{%
    \normalfont B\kern-0.5em{\scshape i\kern-0.25em b}\kern-0.8em\TeX}}}

\setcopyright{acmlicensed}
\copyrightyear{2018}
\acmYear{2018}
\acmDOI{XXXXXXX.XXXXXXX}
\usepackage{array}
\usepackage{longtable}
\usepackage{tabularray}
\usepackage{booktabs}
\usepackage{multirow}
\usepackage{rotating}
\usepackage{booktabs}
\usepackage{adjustbox}
\usepackage{subfigure}
\acmJournal{POMACS}
\acmVolume{37}
\acmNumber{4}
\acmArticle{111}
\acmMonth{8}

\begin{document}

%%
%% The "title" command has an optional parameter,
%% allowing the author to define a "short title" to be used in page headers.
\title{Optimizing Ride-Pooling Revenue: Pricing Strategies and Driver-Traveller Dynamics}

%%
%% The "author" command and its associated commands are used to define
%% the authors and their affiliations.
%% Of note is the shared affiliation of the first two authors, and the
%% "authornote" and "authornotemark" commands
%% used to denote shared contribution to the research.
\author{Usman Akhtar}
%\authornote{U.A is a main author of this paper}
\email{usman.akhtar17@gmail.com}
\orcid{0000-0003-4553-0550}
\author{Farnoud Ghasemi}
\author{Rafal Kucharski}
%\authornotemark[1]
%\email{webmaster@marysville-ohio.com}
%, , , Poland
\affiliation{%
  \institution{Jagiellonian University, Faculty of Mathematics and Computer Science}
  \streetaddress{Prof. Stanisława Łojasiewicza 6}
  \city{Krakow}
  %\state{Ohio}
  \country{Poland}
  \postcode{30-348 }
}

%%
%% The abstract is a short summary of the work to be presented in the
%% article.
\begin{abstract}
  Ride-pooling, to gain momentum, needs to be attractive for all the parties involved. This includes also drivers, who are naturally reluctant to serve pooled rides. This can be controlled by the platform's pricing strategy, which can stimulate drivers to serve pooled rides. Here, we propose an agent-based framework, where drivers serve rides that maximise their utility.
We simulate a series of scenarios in Delft and compare three strategies. Our results show that drivers, when they maximize their profits, earn more than in both the solo-rides and only-pooled rides scenarios. This shows that serving pooled rides can be beneficial as well for drivers, yet typically not all pooled rides are attractive for drivers. The proposed framework may be further applied to propose discriminative pricing in which the full potential of ride-pooling is exploited, with benefits for the platform, travellers, and (which is novel here) to the drivers.
\end{abstract}

%%
%% The code below is generated by the tool at http://dl.acm.org/ccs.cfm.
%% Please copy and paste the code instead of the example below.
%%


%%
%% Keywords. The author(s) should pick words that accurately describe
%% the work being presented. Separate the keywords with commas.

\received{20 February 2007}
\received[revised]{12 March 2009}
\received[accepted]{5 June 2009}

%%
%% This command processes the author and affiliation and title
%% information and builds the first part of the formatted document.
\maketitle

\section{Introduction}
In recent years, ride-hailing service providers, such as “UberPool” and “Lyft Shared", have promoted their ride-pooling business in order to fill empty seats in cars and reduce the monetary costs of rides~\cite{ashkrof2020understanding}. Although ride-pooling has proven to be an attractive new form of transport, it is still fails to reach sufficient demand levels to make it efficient and sustainable~\cite{shen2020modeling}.  In contrast to the traditional transport system, where one car takes at most one rider at a time, the ride-hailing service provider can assign multiple orders of rides to a vehicle simultaneously. Travelers are incentivised to opt for pooled rides rather than solo rides by discounts offered to compensate for discomfort and delay. For travelers, they get a travel fee reduction for sharing rides with others, and for ride-hailing service providers, these sharing rides enable better utilization of the seats, which means more profits. 

\begin{figure}[!t]
    \begin{center}
    \hspace*{-0.70cm} 
       \includegraphics[clip,  trim=0.5cm 0cm 0.3cm 0.2cm, scale=0.95]{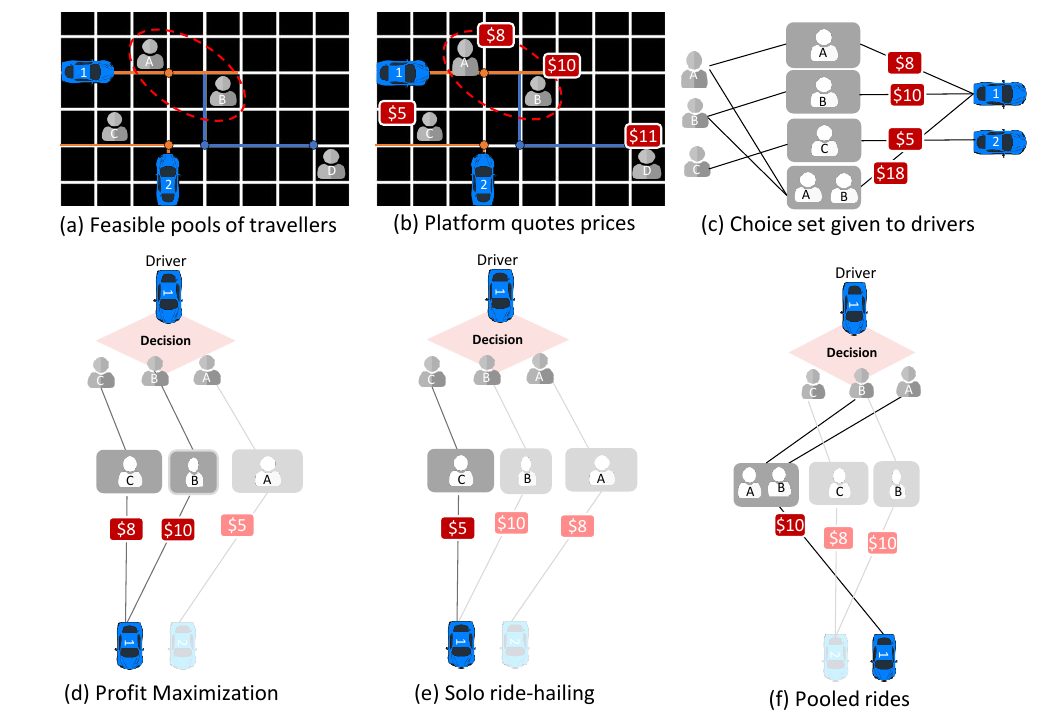}
    \end{center}
     \caption{\small Consider a scenario where the trip requests are submitted to the platform, we first identify attractive pooled options for travellers (a). The platform then determines fares for each ride, depending on its current pricing policy (b). Drivers then are given the choice-set (c), from which they select the optimal ride to serve. In this particular case, driver  selects to serve travellers A and B in a pooled ride with the profit-maximization policy (d) and the solo-ride of traveller (e) under the solo-rides policy (f). The objective is to identify such pricing that maximize the system performance from the perspectives of all the parties involved.}
        \label{fig:IdeaDiagram}%
\end{figure}

Extant opinions gathered from a variety of sources suggest that  Drivers generally dislike UberPool and Lyft shared service. A major source of driver dissatisfaction appears to be monetary compensation~\cite{morris2020drivers}. UberPool paid the driver based on the miles and minutes driven  and there is no compensation for UberPool trips despite the longer pickups and drop-off distances. However, UberPool argues that net income increases driver earnings by offering more total trips~\footnote{https://www.uber.com/us/en/drive/services/shared-rides/}. Another study claims that because pooled rides are typically longer, the driver earns more money from serving pooled rides than UberX trips (which is non shared rides)~\cite{chandar2019drivers}. Nonetheless due to the "lottery" nature of the pooled ride experience, where the driver does not know how much he has to detour they remain unstatisfied and unwilling to pool.

Consequently, the design of the optimal pricing policy becomes an essential business strategy. Many challenges exist to design a sustainable framework while maximizing the overall profit of the platform. The key challenge is to satisfy both types of users in the system e.g., travellers and drivers. However, the majority of the existing work focuses on minimizing  the total travel distance of the drivers.  Ride-pooling platforms use the pricing strategy as a means to control the network's demand (travelers) and supply (drivers) to increase revenue. While ride-pooling companies can have several objectives, it is obvious that revenue maximization always has been one of the main goals. Companies are reluctant to share the detail of their pricing strategies because of their financial situation. Recently, many studies have focused on the dynamic pricing of the solo rides~(\cite{bimpikis2019spatial, chen2020pricing, guda2017strategic}) focus on optimizing ride-sourcing pricing such as determining ride fares and driver wages~(\cite{storch2021incentive, fielbaum2022split, beojonea2022repositioning}). While a variety of studies experimented with optimal pricing for ride-hailing systems with solo rides, pricing for pooled rides is less examined.

The landscape of modern transportation has been significantly influenced by the emergence of ride-sharing platforms. Consider a scenario in which trip requests are submitted to the platform, as shown in Fig.~\ref{fig:IdeaDiagram}. In this framework, the process unfolds systematically: initially, identifying appealing pooled options for travelers. Subsequently, the platform undertakes the task of determining the fares for each ride, a decision upon the prevailing pricing strategy. Drivers are then presented with a choice-set comprising various ride options, from which they strategically select the most profitable ride to serve.

The ride-pooling platform will receive the request and find the driver that offers a solo or pooled rides.   We also assume that drivers are rational utility maximisers and model their choice process. For travelers, we assume that they are interested in pooling, only if the induced detour and delays are compensated with the fare discount related to their perceptions (value of time and willingness to share). To offer a driver a choice set of requests based on the price to serve the rides.  The objective of the platform is to maximize revenue by finding the optimal rides to serve the request. In particular, the platform will offer an optimal ride choice to offer and the corresponding driver. We proposed three pricing policies for the given scenario of solo ride-hailing, forced-pooling, and third intermediate profit maximization. In the private ride-hailing scenario, we assume that there are only single rides (like UberX) and that no pooled rides are served. In forced pooling, we only consider pooled rides, and solo rides are excluded from the simulation. The travellers who cannot be pooled with the other cannot be served. In profit-maximization, we assume that drivers maximize their profit based on the information provided (costs and revenue). In other words, drivers always choose the ride with the highest profit in the requests choice sets. And the profit depends on the platform pricing policy; here we assume the flat commission rate.

So far, the pricing policies with the objective to induce effective pooling were not proposed.  We experiment with a medium-sized city (Delft, The Netherlands) and observed how varying pricing policies affect the system performance. Due to multiple individual agents involved in the system, proposing a complete and meaningful set of indicators (KPIs) is crucial to understand the non-trivial trade-offs and disentangle perspectives of competing parties. With such an approach, we can evaluate the number of various pricing policies with the objective to align the ride-pooling satisfaction among the three parties involved: travellers, platforms, and drivers. This allows a study to investigate the social implications of a profit-maximizing pricing strategy, and consequently the need for pricing regulations in the effective ride-pooling market. We now summarize the key contributions of this paper.

\begin{itemize}
  \item We systematically evaluate and compare three distinct pricing strategies: profit maximization, pooled rides, and solo rides, providing a detailed analysis of their impact on service rates, revenue generation, platform earnings, waiting times, and occupancy levels. 

  \item Through extensive simulations and experiments, we demonstrate that the profit maximization strategy stands out in terms of both service rate efficiency and revenue generation. It engages drivers effectively and maximizes the platform's overall revenue, showcasing its potential for optimizing resource utilization. 

  % \item  We emphasize the importance of an even distribution of earnings among drivers, particularly highlighting the pooled ride strategy's ability to achieve a more equitable revenue distribution compared to the profit maximization approach. This finding underscores the significance of considering driver revenue and sustainability in pricing policies. 
  
  \item By identifying the strengths and weaknesses of various pricing strategies, we pave the way for future research in developing hybrid pricing models that leverage the advantages of profit maximization and pooled rides, aiming to strike a balance between platform, drivers, and the travellers. 
  
  %We conduct experiments to investigate the performance of the proposed profit maximization strategy and compare it with them with a state-of-the-art approach.  

  %Theimportant notations and symbols in the rest of this paper are sum-marized in Table 1 .
  
 % \item An item with an \textsc{equation}:

\end{itemize}

The remainder of this paper is organized as follows: Section 2: Related Work explores existing literature and research related to ride-pooling and pricing strategies. This section sets the foundation for our research by presenting an overview of the current state of the field and identifying areas that require further investigation. In Section 3, we describe the methodology in our research, an agent-based framework, and the simulations conducted to analyze the impact of different pricing strategies. In Section 4, we present the results of our experiments comparing three distinct pricing strategies: profit maximization, pooled rides, and solo rides. We show case findings related to service rates, revenue generation, platform earnings, waiting times, and occupancy levels. Finally, in terms of discussion and future presented in Sections 5 and 6, we engage in a detailed discussion of the results and their implications. We also highlight the contributions of this research and discuss potential future directions that could further optimize the benefits of ride-pooling platforms.

\section{Related Work}

%Our work is related to the burgeoning literature that explores the pricing strategy of ride-hailing. %The literature on pricing strategies in ride-hailing can be broadly divided into two categories: (i) pricing strategies for solo rides and (ii) pricing strategies for pooled rides. 

%Pricing strategies for solo rides in ride-pooling are based on several factors such as distance, travel time, and detours. 

%%% Shared ride pool pricing 
The burgeoning literature is devoted to discussing pricing strategies for ride-pooling. Recent studies have shown that expected driving income has a great influence on driver participation~\cite{fang2017prices} in ride-haling, and cost also plays an important role in its participation~\cite{asghari2018adapt}. On the other hand, empirical studies on travellers show that abrupt increases in Uber surge pricing dramatically reduce demand for travelLers~\cite{chen2015peeking}.

Estimating the demand elasticity and maximizing expected profit is the main focus of pricing strategies for ride-pooling. Some literature attempts to find the optimal solution for a specific objective, such as to capture the temporal elasticity of the demand~\cite{sayarshad2015scalable, qian2017time}. ~\cite{sayarshad2015scalable} proposed a non-myopic pricing method for a pooled ride using multi-server queues that estimate future costs and profits in relation to the quality of the service and price. Inspired by non-myopic pricing, ~\cite{qian2017time} develops a time-of-day pricing scheme to maximize the profit of ride-hailing; a strict pricing scheme should be considered for heterogeneity in the demand. ~\cite{he2018pricing} proposed a penalty compensation strategy to cancel pooled rides to maximize profit for drivers. 

In terms of passenger preference, some have applied the Multinomial Logit (MNL) model to estimate the in an agent-based framework presented in~\cite{chen2016management}. They also investigated the trade-offs between revenue and ride service under different pricing strategies such as distance-based pricing and origin-based pricing. ~\cite{qiu2017dynamic} 
 proposed a framework to solve the problem of profit-maximization in shared private mobility, also considering the preference of travellers, the estimation of demand, and the congestion of the network.

However, these approaches do not consider the impact of the pricing strategies on driver acceptance probability, as they only consider the fixed flat rate pricing scheme. Furthermore, they assume that drivers will always accept assigned rides, as described by ~\cite{ashkrof2020understanding}, drivers are free to decide acceptance of the ride. On the other hand, dynamic pricing, also known as surge pricing, also depends on the supply-demand; Compared with the flat-rate pricing schemes, surge pricing identifies the payment of the driver based on the supply-demand conditions of the region. For profit maximization, surge pricing is designed to relocate drivers so that the gap between supply and demand is mitigated. Some pricing strategies are reward-based, as proposed by the ~\cite{yang2020integrated} proposed scheme to allow travelers to pay an additional amount during peak hours. The author concluded that reward-based approaches improve the utility of travellers, driver income, and platform revenue. 

\begin{table}[H]
    \caption{Summary of Related Work}
    \label{table:relatedwork}
    \begin{tabular}{@{}p{0.5cm}p{3.5cm}p{3.5cm}p{2cm}p{1.5cm}p{1.5cm}p{1.5cm}@{}}
        \toprule
        \textbf{Work} & \textbf{Advantages} & \textbf{Disadvantages} & \textbf{Method} & \textbf{Solo-Rides} & \textbf{Pooled-Rides} & \textbf{Profit Maximization} \\
        \midrule
        \cite{sayarshad2015scalable}  & Uses non-myopic pricing based on routing policy function    & Does not consider pricing of pooled rides, only dispatch, routing, and pricing instructions from a central authority & Non-myopic pricing component & $\checkmark$ & -- & -- \\ 
        \cite{qian2017time} & Proposes dynamic fare pricing technique to minimize fares for passengers and maximize profit for platform  & Dynamic fare pricing sometimes leads to road congestion, promotes private transportation, consumes more energy, and leads to more emissions  & Mathematical optimization based technique & --  & $\checkmark$ & -- \\  
        \cite{he2018pricing}  & Proposes pricing and penalty strategy based on reward for shorter waiting times   & Determining proper reward and cancellation causes discomfort for travelers  & Equilibrium model & $\checkmark$ & -- & -- \\  
        \cite{chen2016management}  & Proposes pricing strategies to balance available supply with anticipated trip demand and decrease average wait time   & Trade-offs exist favoring higher revenue-to-cost ratio with high value of travel time  & Agent-based framework & $\checkmark$ & --  & -- \\  
        \cite{qiu2017dynamic}  & Proposes dynamic pricing schemes to adjust supply and demand mismatch based on multinomial logit model & Traveler decision process in multimodal market not considered & Multinomial logit model & -- & -- & $\checkmark$ \\  
        \cite{zha2018geometric} & Develops model to investigate effects of spatial pricing on ride-sourcing markets, matching customers with nearby drivers   & Only focuses on demand and impact of ride-sourcing services without specific investigation on ride-splitting & Discrete time geometric matching framework & -- & $\checkmark$  & -- \\
        \cite{yang2020integrated} & Proposes reward-based approach to address limitations and reduce potential negative impacts of surge pricing   & The reward-based idea remains unclear how much of this potential can be achieved by profit-maximizing platform & Integrated reward scheme & --  & $\checkmark$  & -- \\
        Our Work    & Proposes profit maximization strategy to find optimal solutions for travelers, drivers, and platform, validates various scenarios for solo-ride, forced pooling, and profit-maximization & High service rates achieved when constraints are relaxed or there is high resource availability  & Profit-Maximization Framework & $\checkmark$  & $\checkmark$  &  $\checkmark$ \\
        \bottomrule
    \end{tabular}
\end{table}

\section{Methodology}

Our methodology is structured to analyze the complex interplay between travellers, drivers, pricing strategies, and system performance. The core elements in our approach encompass the choice-sets given to the driver through a dynamic pricing mechanism facilitated by the platform. The platform sets the pricing for the travellers (fare per kilometer) and the discount offered for pooling (relative to the solo-ride price). We also identify the attractive pooled ride combinations with the utility-driven ExMAS~(\cite{kucharski2020exact}) algorithm which matches trips to attractive shared rides and an agent-based two-sided mobility simulator MaaSSim~(\cite{kucharski2022simulating}). The driver, receiving the set of unserved requests, assesses his costs (distance and time to pick-up and to serve the demand) and the benefits (fare). The fare, in turn, is determined by the platform, which in general is free to decide how much the driver will receive for serving the given demand. Typically, this is a flat commission rate, yet with the proposed framework we may explore with more advanced pricing strategies. Finally, we let drivers choose their optimal trips to serve. We look at individual travellers with their waiting times, travel times, fares paid, and general satisfaction. We also look at the drivers with their idle time, distance travelled, costs, revenues, and the total collected commission fees. %Finally, we evelauated at the system performance, its efficiency, welfare created, and total miles travellers. Such multidimensional view allows for a comprehensive overview of the system performance.

%This research is based on an existing algorithm, ExMAS~(\cite{kucharski2020exact}), which matches trips to attractive shared rides and an agent-based two-sided mobility simulator MaaSSim~(\cite{kucharski2022simulating}). First, we briefly introduce them followed by the introduced pricing methods and driver choice behavior. An overview of the proposed methodology is shown in Fig.~\ref{fig:IdeaDiagram}.  

\subsection{Matching travellers to attractive pooled rides (ExMAS)}

To represent ride-pooling, we use the utility-driven ExMAS algorithm that works offline and identifies any feasible pools of travellers. In its utility-driven formulation, ExMAS identifies rides that are attractive to all travellers, that is, for which utility (composed of travel time, cost and discomfort) of the pooling ride is greater than utility for the solo-ride. The attractiveness is controlled on one side with the behavioural parameters of the traveller (value of time and penalty for sharing), on the other with the spatiotemporal similarity of trips, and most notably, via the discount offered by the platform provider for sharing. Here, we assume that the costs of ride-pooling are at least compensated with a discounted trip fare. ExMAS uses efficient graph searches to narrow the travellers' search space, thereby determining the exact solution of feasible/attractive shared rides. ExMAS precomputes a wide set of any feasible pooled rides for the demand set (trip requests). The rides in this set may be pooled (where two or more travellers ride together) or private. Each traveller is assigned at least to the private ride and can be a member of multiple pooled rides (depending on the feasibility of pooling with others). The details of the ExMAS algorithm are given in \cite{kucharski2020exact}.

\subsection{Ride-Hailing Operations with Agent-Based simulator (MaaSSim)}

We integrate our pricing strategies into MaaSSim, an agent-based event simulator for mobility-on-demand operations. The agent-based nature of this model allows us to capture heterogeneity in ride-pooling earnings between travellers, drivers, and platforms. We simulate the two-sided mobility market with MaaSSim, where individual drivers serve incoming requests from travellers in real-time. 
Both travellers and drivers operate on a detailed network graph and can make individual decisions.  Thanks to the utility-driven nature of the ExMAS algorithm, all the proposed pooled services are, by definition, attractive to travellers; thus, we may limit the decision-making process in this study to drivers only.

\subsection{Drivers' ride-pooling decision process}

For a given demand dataset, we run an event-based microscopic simulation where each ride is requested at the moment of departure. Requesting a ride triggers the matching process for drivers. The nearest idle driver receives a request and makes a decision; his choice set is composed of all feasible rides. Typically, the rider may choose to serve the first requesting travellers alone or in one or more pooled options. 

We assume that the driver makes a choice that maximizes her utility. In the deterministic case, the option with the greatest profit is chosen; in the probabilistic case, the multinomial logit model is applied. The utility is expressed in terms of revenues (transferred from travellers via the platform to the driver according to the pricing policy), from which we deduct the costs (value of time and operating costs). In line with recent empirical findings, we optionally penalize the pooled ride with a multiplier. The multiplier may be homogeneously distributed across the population or (more realistically) heterogeneous and varying with the pooling aptitude of individual drivers.

The fare paid by the traveller $i$ is distance-based $l_i$ and discounted by $\lambda$ if the ride $r$ is pooled:

\begin{equation}
     f_{i,r} = \begin{cases}
      \beta_c  l_i, \& \text{if}\ \text{private ride} \\
      \beta_c (1-\lambda) l_i, \& \text{pooled ride}
    \end{cases}
\end{equation}

The total fare collected by the platform to serve the ride is the sum of fares paid by the travellers:

\begin{equation}
    F_r = \sum_{i \in r} f_{i,r}
\end{equation}

driver receives the total fare minus the commission $T_r$:

\begin{equation}
    R_r = (1-T_r)F_r 
\end{equation}

and by reducing the operating costs of serving the ride $C_r$ we obtain his profit:

\begin{equation}
    I_r = R_r - C_r
\end{equation}

The costs of operating the ride are expressed as the sum of the cost per km $\beta_l$ multiplied by the duration of the ride and the distance to pick up $l^p_{i,d}$ and the value-of-time $\beta_t$ multiplied by the duration of the trip $t_i$ and the time to pick up $t^p_{i,d}$. Note that while all the components so far were trip-specific, the pick-up distances depend on the driver's current position:

\begin{equation}
    C_r = \beta_l (l^p_{i,d} + l_i) + \beta_t (t^p_{i,d} + t_i)
\end{equation}

The utility of the ride is then the profit associated with a ride multiplied by the multiplier $\beta_{r,i}$ being 1 for private rides and presumably lower for pooled rides for drivers with reluctance to serve the shared ride. The probability of selecting the ride $r$ from the set of unserved rides $\mathbf{r}$ by the driver $i$ is then:

\begin{equation}
    P_{r,i} = \text{Pr} \left( \beta_{r,i}I_r = \max_{r' \in \mathbf{r}} \beta_{r',i}I_{r'} \right)
\end{equation} \label{eq:prob}

which may be deterministic or probabilistic depending on the experimental setting.

\subsection{Pricing policies}

We examine the system performance in three pricing strategies: private ride-hailing, forced pooling, and profit maximization. %In private and forced pooling, the platform matches drivers with the closest ride. 
%\subsection{Pricing policies}

\begin{itemize}
  \item {In the private ride-hailing scenario, we assume that there are only single rides (like UberX) and no pooled rides in this scenario.}
  \item {In the forced pooling scenario, we  consider having only pooled rides and solo rides are excluded from the simulation. The travellers who cannot be pooled with others, are not served.}
\item {In the intermediate scenario: profit maximisation, we assume that drivers maximize their profit based on the information provided (costs and revenues). In other words, drivers always choose the ride with the highest profit in the choice sets, which can be either pooled or solo rides.}
\end{itemize}

  \section{Experiments}

  We evaluate the efficacy of the proposed pricing strategies within a MaaSSim (simulator) utilizing the road network of Delft, represented as a directed graph constructed via OSMNnx~\cite{boeing2017osmnx}. Our comparison involves contrasting our approach with two alternative pricing strategies: private ride-hailing and forced pooling.

  \subsection{Experimental Setup}

In the experimental setup, the investigation is conducted using MaaSSim~\cite{kucharski2022simulating}, a Python-based open-source agent-based simulator, with a total simulation time of 4 hours ($T_{simTime} = 4$ hours). Different scenarios are explored, varying the number of travelers from 100 to 500 and potential drivers from 5 to 50. Benchmark scenarios are considered, where a platform offers a constant discount level to all travellers matched in shared trips, accompanied by a commission rate of 0.25 and a fare of 1.5€/km. The operational costs for drivers, covering fuel, depreciation, etc., amount to 0.5€/km, deducted from the revenues to calculate the drivers' profit. The parameter values for the experiments, detailed in Table~\ref{Expparameter}, encompass various aspects such as simulation time, network speed, city representation, demand parameters (maximum requests per hour and traveller patience), supply parameters (maximum number of drivers and driver-specific parameter $\beta_{i}$), and platform-related values (commission rate and constant discount). These values are fundamental for understanding the experimental conditions and outcomes within the MaaSSim simulator.

\begin{table}[h!]
\centering
\caption{Parameter values in the experiment}
\label{Expparameter}
\begin{tabular}{@{}llll@{}}
\toprule
\textbf{Type} & \textbf{Parameter} & \textbf{Value} & \textbf{Unit} \\
\midrule
Simulation Time & $T_{\text{simTime}}$ & 4 & h \\
Network & Speed & 36 & km/h \\
& City & Delft & graph \\
Demand & Travellers & up to 400 & requests/h \\
& Patience & 5 & minutes \\
Supply & Driver & up to 20 & drivers \\
Platform & $\lambda$ & 0.25 & discount \\
\bottomrule
\end{tabular}
\end{table}

In our experimental framework, we systematically vary key parameters to measure essential metrics reflecting the efficacy of our pricing strategies in ride-sharing services. The Service Rate (SR) is a pivotal metric, quantifying the percentage of fulfilled requests achieved through the applied pricing strategy. Revenue metrics, which include profit maximization, pooled and solo rides, as well as total revenue  by the platform, are crucial aspects examined in our experiment. These metrics are specifically tailored for both pooled and private rides. Additionally, the article delves into understanding the satisfaction of travellers through waiting time ($w$), and occupancy ($u$), a composite measure that involves travelers' time, driving time, and discomfort, designed differently for pooled and private rides. These metrics provide valuable information to optimize ride-sharing services and improve the overall experience for  travellers, platform and drivers. 

\begin{itemize}
    \item Service Rate (SR): Percentage of requests that were fulfilled by the pricing strategy.
    \item Revenue Metrics for drivers and platform:
    \begin{itemize}
        \item Revenue generated by the pricing strategy, including profit maximization, pooled and solo rides.
        \item Total revenue earned by the platform.
    \end{itemize}
      \item  Demand Satisfaction:
     \begin{itemize}
        \item waiting time $w$
        \item travel time $t$
        \item Occupancy $u$ (expressed with the ExMAS ride-pooling formula, composed of travellers time, driving time, and discomfort - different if the ride was pooled or private).
    \end{itemize} 
   % \item Total travel time and distance - sum of distances and times of all the drivers.
    \end{itemize}

    These indicators comprehensively evaluate the performance of a ride-sharing service, encompassing the level of service (LoS) for travelers, drivers, and service providers.

\subsubsection{Service Rate and Pooling}

\begin{figure}[!htb]
    \centering
    \subfigure[Service Rates with 20-Minutes Patience Time]{%
        \includegraphics[clip, scale=0.50]{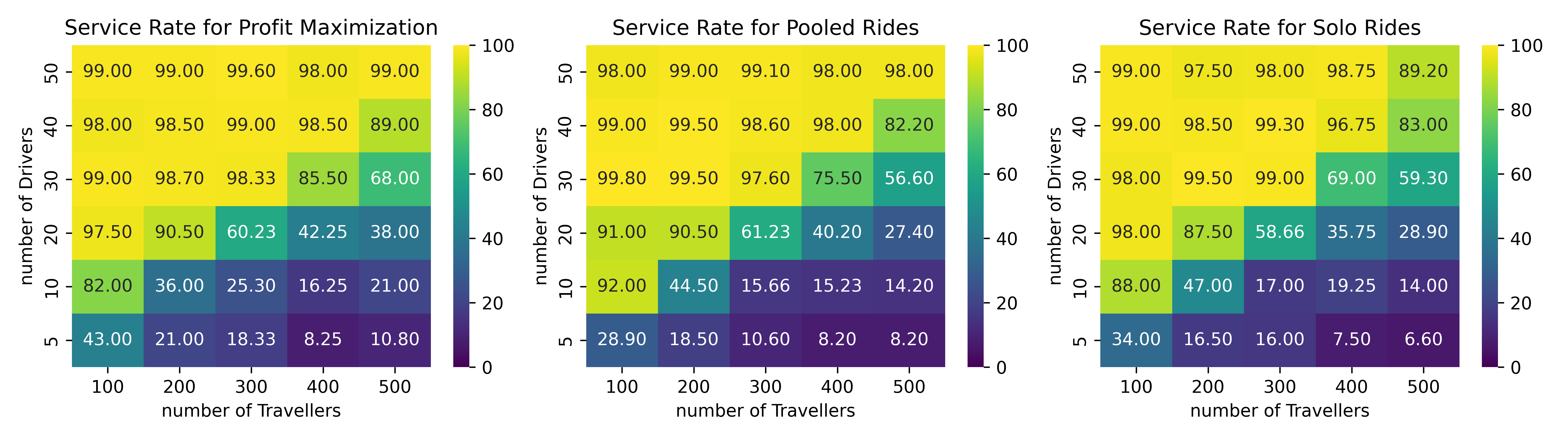}
        \label{subfig:unlimited_patience}
    }\hfill
    \subfigure[Service Rates with 3-Minute Patience Time]{%
        \includegraphics[clip, trim=0.15cm 0.25cm 0.25cm 0.25cm, scale=0.47]{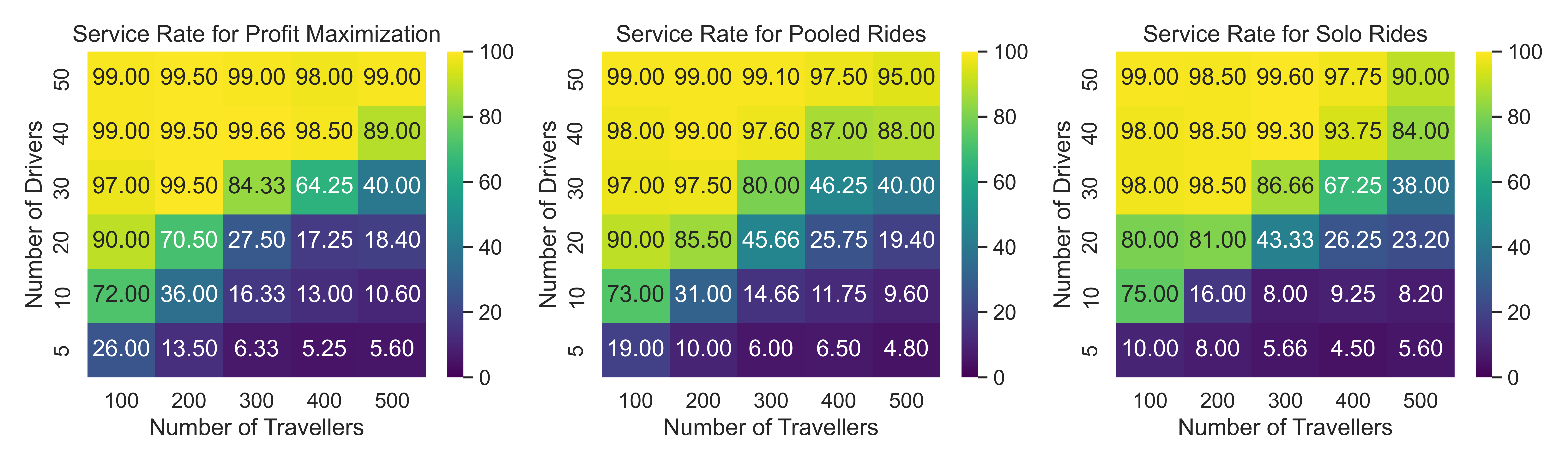}
        \label{subfig:3min_patience}
    }
    \caption{We analyze the service rates (\% of requests serviced) for three different strategies: profit maximization, pooled rides, and solo rides.   On a grid of supply and demand Figure (a) shows the service rates for profit maximization, pooled rides, and solo rides with twenty minutes of patience time constraint. Profit maximization strategy efficiently utilizes available resources, resulting in a higher percentage of incoming ride requests being serviced.
    Under the constraint of a 3-minute patience time as shown in Figure (b), we have examined the service rates (\% of requests serviced). Our analysis reveals that the profit maximization strategy consistently outperforms the other two approaches.}
    \label{fig:service_rate_comparison}
\end{figure}

We evaluated the service rates associated with three different strategies: Profit maximization, single and pooled-rides with the objective of efficiently matching more ride requests to available drivers and achieving a higher service rate. The results, presented in Figure~\ref{subfig:unlimited_patience}, highlight the significant success of the profit maximization strategy, showing up to 10\% increase in the service rate compared to other strategies while maintaining a waiting time of 20 minutes. The higher service request depends on the availability of the driver, and profit maximization engage drivers more effectively hence able to serve more requests with higher service rate.  This profit-maximization approach proved to be highly effective, resulting in increased service rates, platform earnings, and overall revenue, outweighing the revenue lost from rejected requests.

In another analysis, we investigate by varying the waiting-time constraints on service rates. Our findings indicate a significant drop in service rates under tight waiting time constraints, resulting in the rejection of a notable proportion of ride requests. The effect of varying the maximum waiting time is 3 minutes as shown in Fig.~\ref{subfig:3min_patience}. Interestingly, as the waiting time for served requests increases, there is a corresponding improvement in the overall service rate(\%). Longer waiting times allow for a higher number of requests to reach the source location of the ride. This, in turn, increases the likelihood that drivers fulfilling these requests.

   We observed that the effectiveness of the solo and pooled ride strategies decreases under the tight constraint of 6 minutes of waiting time. On the other hand, Profit maximization assigns more requests under tight constraints of waiting time. However, it is worth noting that as we sharpen the the waiting time, the service rate exhibited an upward trend. This suggests that longer waiting times have a positive effect on service rate, highlighting the potential benefits of providing more flexibility in waiting-time constraints. Our experiment underscores the intricate relationship between waiting time constraints and service rates in ride-pooling scenarios. These findings emphasize the need to carefully consider waiting-time policies in conjunction with different pricing strategies to ensure optimal service rates (\%) and travellers satisfaction.
   
In conclusion, our comprehensive analysis of service rates for profit maximization, pooled rides, and solo rides under varying waiting time constraints has provided valuable insight into the dynamics of ride-pooling strategies. The profit maximization strategy consistently exhibited superior performance, efficiently utilizing available resources, and outperforming other approaches, whether subjected to a 20-minute or 3-minute waiting time constraint. Additionally, our exploration of the impact of waiting time constraints on service rates highlighted the relationship between waiting time and the effectiveness of different ride-pooling strategies. As we consider the potential benefits of longer waiting times on service rates, it underscores the importance of carefully aligning waiting-time policies with pricing strategies to optimize service rates.

\subsubsection{Comparing Revenue of the Pricing Strategies}
\begin{figure}[!htb]
    \begin{center}
    \hspace*{-0.70cm} 
       \includegraphics[clip, trim=0.15cm 0.25cm 0.25cm 0.25cm, scale=0.50]{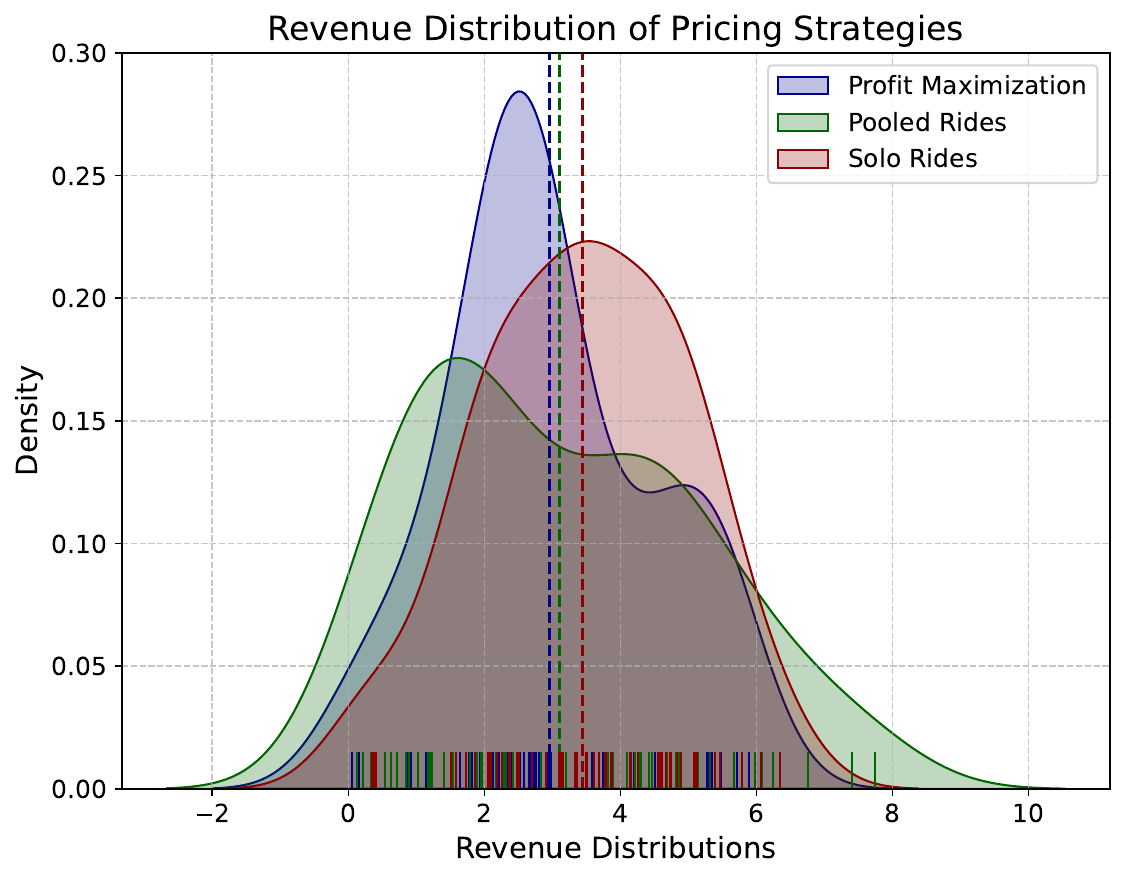}
    \end{center}
     \caption{\small The diagram shows the revenue distribution of profit maximization, pooled rides.  These rides are associated with a total of five hundred travelers and fifty drivers. The x-axis and the y-axis shows the  revenue distribution generated by each pricing strategy. Profit maximization generate highest revenue for the individual drivers but the the revenue is unevenly distributed.  Pooled  rides generate the least revenue overall and are also the most unevenly distributed. Solo rides generate less overall revenue, but are more evenly distributed, which can be beneficial to drivers.}
        \label{fig:RevenueDist}%
\end{figure}

In this experiment, we conducted a detailed comparison of revenue (that is, total income or earnings generated through different pricing strategies employed in a ride-sharing platform), as illustrated in Figure \ref{fig:RevenueDist}. The analysis focused on the revenue distribution among 50 drivers to evaluate the profitability of existing strategies: profit maximization, pooled rides, and solo rides.  Contrary to initial observations, our findings reveal that profit maximization does not uniformly generate higher revenue across individual drivers when compared to pooled rides. Rather, it is evident that profit maximization yields higher total revenue, but this revenue distribution is characterized by significant unevenness, with a few drivers contributing substantially while others contributing minimally. However, the solo rides strategy, aimed at maximizing revenue per individual ride, not only generates the least overall revenue but also exhibits the most uneven distribution among participating drivers. It is crucial to emphasize that the observed revenue distribution highlights a limitation of the solo-ride strategy, where only a select few drivers contribute significantly to revenue generation, leaving a substantial number of drivers with minimal revenue. Focusing on maximizing revenue per individual ride in solo rides may result in fewer drivers actively engaged, contributing to the observed uneven revenue distribution. 

On the contrary, the pooled ride pricing strategy demonstrates a more uniform revenue distribution. This outcome is attributed to the strategy's ability to involve more drivers in revenue generation.  In conclusion, our analysis sheds light on the intricate dynamics of revenue distribution among drivers under different pricing strategies. The findings underscore the importance of balancing total revenue generation with consideration of the equitable distribution of earnings among participating drivers, offering valuable information to optimize pricing strategies on ride-sharing platforms.

\subsubsection{Comparing Revenue earned by the Platform}
\begin{figure}[!htb]
    \begin{center}
    \hspace*{-0.85cm} 
       \includegraphics[clip, trim=0.15cm 0.25cm 0.25cm 0.25cm, scale=0.55]{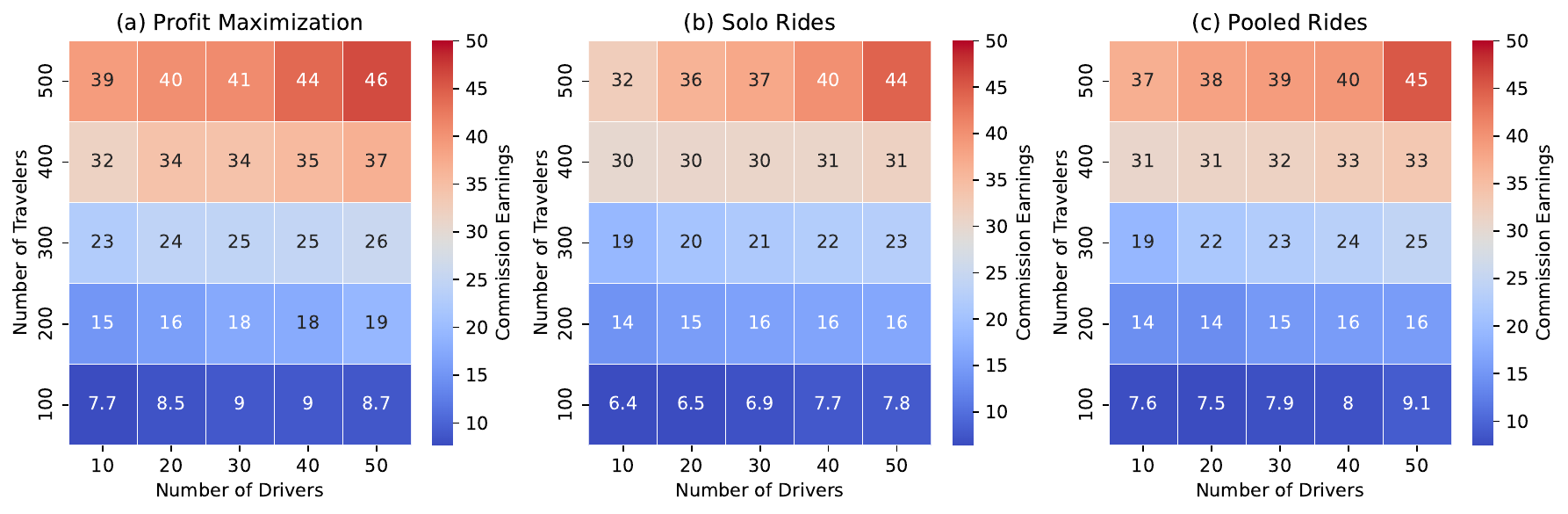}
    \end{center}
     \caption{\small The  diagrams illustrate the commission earnings garnered by a ride-sharing platform across three distinct pricing strategies: profit maximization, pooled rides, and solo rides. On the x-axis, the number of drivers adopting each pricing strategy is represented, while the y-axis portrays the platform's cumulative commission earnings corresponding to each strategy.}
    \label{fig:COMMISSION_distribution_all}%
\end{figure}

In this analysis, we examine the revenue earned by the platform under different pricing strategies: profit maximization, pooled rides, and solo rides. As depicted in Figure~\ref{fig:COMMISSION_distribution_all}(a) the platform charges the highest possible price for each ride, resulting in the highest revenue. This is because the platform is trying to maximize its own profit. However, with the increase of the drivers, the revenue also increases, but much slower. The platform has to pay its drivers, which means that the extra money it makes from each additional driver is not as much. The results of the experiment show that the profit maximization strategy is effective in maximizing the revenue earned by the platform. However, it is important to note that the optimal number of travellers and drivers will vary depending on the specific situation. For example, the optimal number of travelers and drivers may be different if the platform is operating in much smaller and crowded cities.

Figure~\ref{fig:COMMISSION_distribution_all}(b) visualizes how the combination of drivers and travelers affects the platform revenue from solo rides. As the number of drivers increases, the platform can potentially handle more rides and accommodate more travellers. This could lead to an increase in total commission earned, as more rides can be completed and more commissions collected. However, it is important to note that the profit generated by the platform will also depend on the fare charged per ride. For solo rides, if the fare is high, the platform will generate more profit, even if the number of travellers and drivers is lower.

Figure~\ref{fig:COMMISSION_distribution_all}(c) 
shows the platform earnings for a pooled ride as illustrated. As travellers increases, the platform has more opportunities to generate revenue through commissions from pooled rides.  In pooled rides, various factors influence the overall profitability of the platform, including operational costs, driver compensation, travel cancellations and travel distances. We can deduce that the profit maximization strategy yields the most income for the platform, yet it also has the highest likelihood of turning away drivers and riders. The pooled rides strategy produces less commission for the platform, but is more sustainable and could be more favored by drivers and riders. 

\subsubsection{Comparison of Traveller Waiting Times Across Ride-Sharing Strategies}

In this study, we conducted an analysis of the waiting times experienced by travellers with ride-sourcing requests. The waiting time encompasses the time between making a ride request and receiving the assignment, as well as the time it takes for the assigned driver to reach the pick-up location. In particular, the waiting time for travellers in pooled rides is significantly longer than that for travelers in private rides.

\begin{figure}[!htb]
    \begin{center}
    \hspace*{-0.70cm} 
       \includegraphics[clip, trim=0.15cm 0.25cm 0.25cm 0.25cm, scale=0.60]{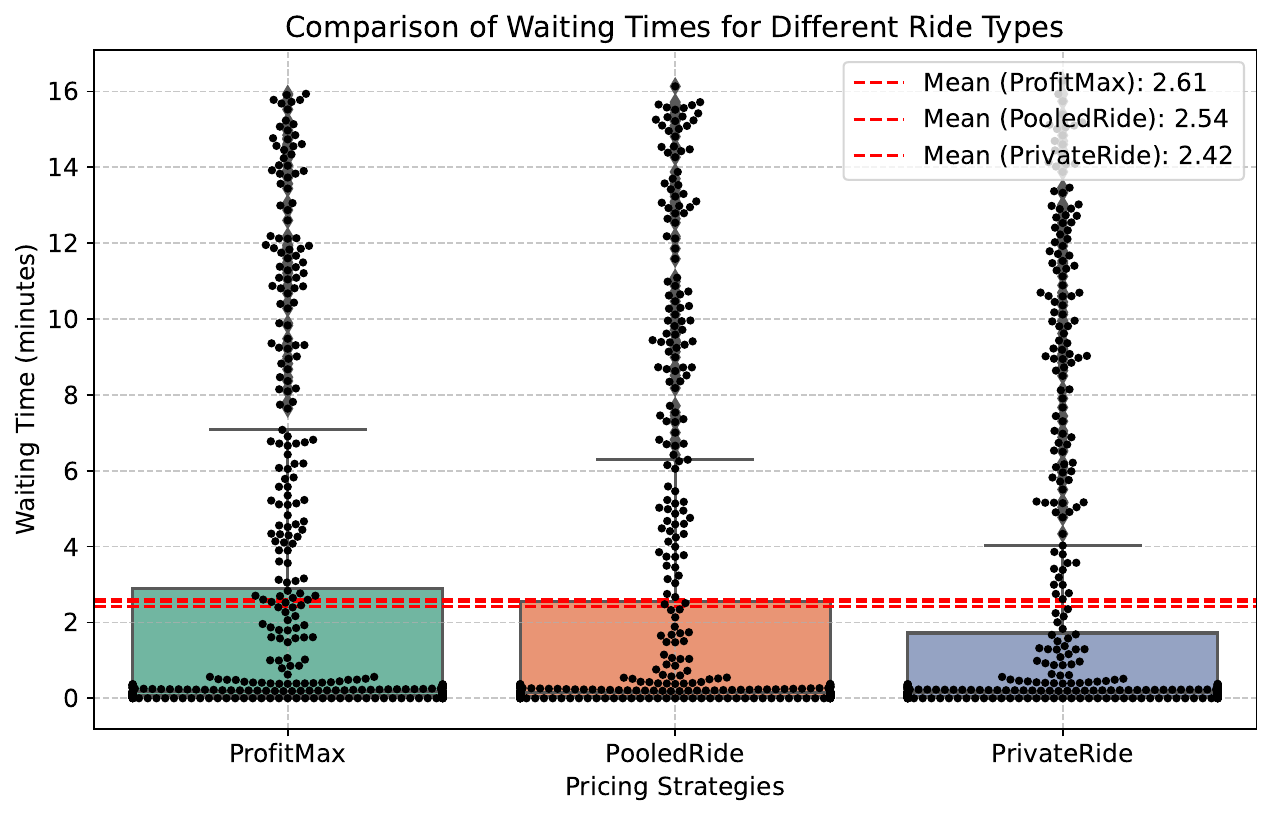}
    \end{center}
     \caption{\small An experiment was conducted involving 500 travellers and 50 drivers to analyze waiting times based on different pricing strategies. Among the strategies, the profit maximization approach exhibited the longest waiting times, followed by the pooled rides strategy, and finally, the solo rides strategy.}
        \label{fig:WaitingTime}%
\end{figure}

In Figure~\ref{fig:WaitingTime}, we present an examination of waiting times for travellers under profit maximization, pooled rides, and solo rides. Specifically, the profit maximization strategy exhibits the longest waiting time compared to solo and pooled rides. The mean waiting time for the profit maximization strategy is 2.61 minutes, whereas the mean waiting time for the private ride strategy is 2.42 minutes. On average, a traveller opting for a pooled ride experiences a 19-second longer wait than a traveler choosing a private ride. This discrepancy can be attributed to the platform's emphasis on profit maximization rather than minimizing waiting times for travelers.

Under the solo rides strategy, waiting times fall between those of pooled and private rides. Travelers are matched only with others going in the same direction, without the obligation to share a ride. The mean waiting time for a pooled ride is 2.54 minutes, slightly more than the 2.42 minutes for a private ride, resulting in a difference of over 12 seconds in favor of private rides. Several factors contribute to this difference, including the additional time required for multiple pick-ups and drop-offs in pooled rides and the higher popularity of pooled rides, leading to increased demand and, subsequently, longer wait times.

\subsubsection{Occupancy Analysis of Ride-Sharing Strategies}

\begin{figure}[!htb]
    \begin{center}
    \hspace*{-0.70cm} 
       \includegraphics[clip, trim=0.15cm 0.25cm 0.25cm 0.25cm, scale=0.50]{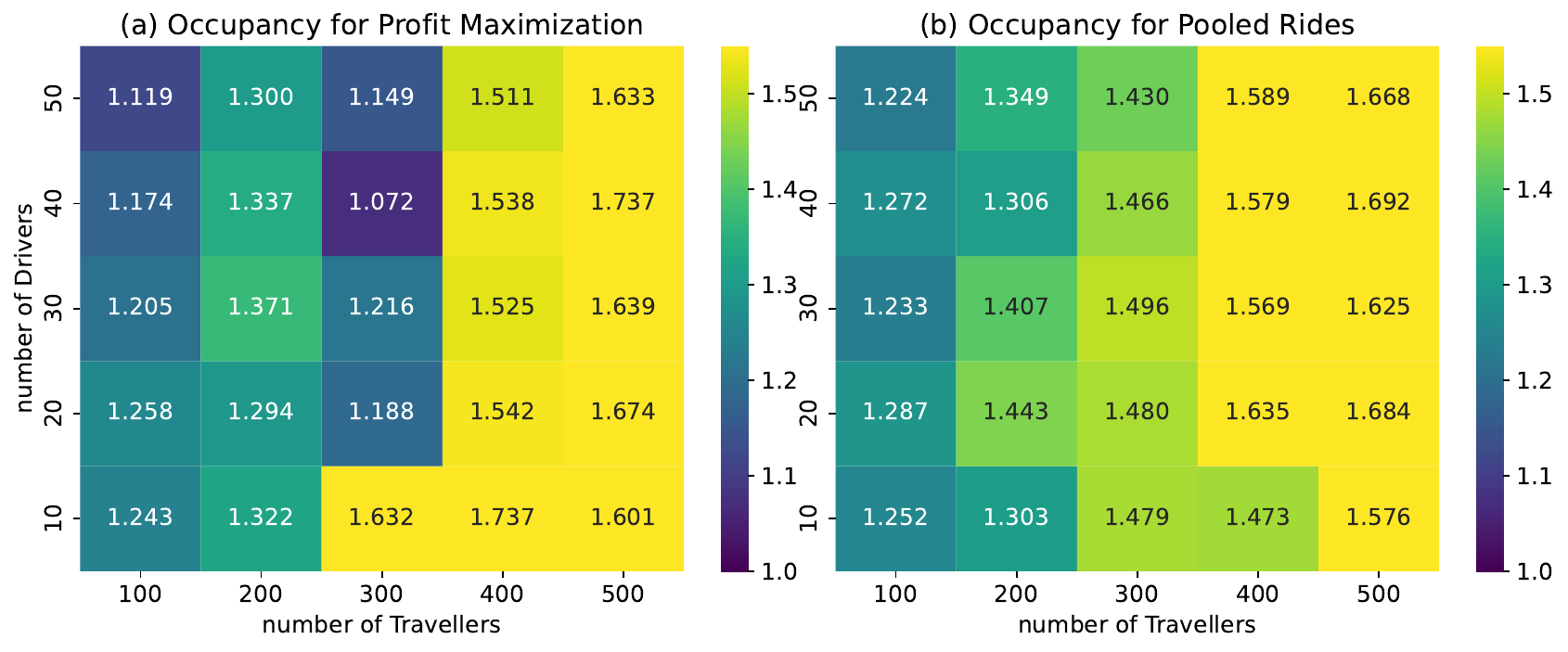}
    \end{center}
     \caption{\small The figure illustrates the occupancy rates for three ride-sharing strategies: profit maximization, solo rides, and pooled rides. Occupancy is calculated as the ratio of traveller travel hours to driver driving hours, providing information on resource utilization and system efficiency.}
        \label{fig:Occupancy}%
\end{figure}

In this analysis, we evaluated the occupancy of both the profit maximization strategy and the pooled rides, calculated as the total travel time of the travelers divided by the total driving time. Occupancy is a measure of the utilization of available driving resources in terms of time spent transporting travellers. Figure~\ref{fig:Occupancy}(a) depicts the occupancy for profit maximization, showcasing a lower occupancy when the number of drivers and travelers is limited. However, with an increase in the number of drivers, the occupancy improves, indicating more efficient utilization of resources.
As illustrated in Figure~\ref{fig:Occupancy}(b) for pooled rides, occupancy values are generally lower, especially when the number of drivers is higher and the number of travelers is lower. This suggests that the pooled ride strategy may prioritize efficiency in ride-sharing over achieving consistently high occupancy. Unlike strategies that prioritize profit or shared rides, which tend to result in higher occupancy levels in specific scenarios, pooled rides may focus more on passenger comfort. Algorithms for pooled rides might accept a lower occupancy rate to minimize rider wait times without emphasizing maximizing occupancy.

\section{Discussion}

In this simulation, we study  the impact of different pricing strategies on the service rate, revenue, platform earnings, and waiting times of ride-sharing platforms. The results showed that the profit maximization strategy resulted in a higher service rate in terms of requests and a higher overall revenue generation for both the drivers and the platform. However, it also led to a more uneven distribution of revenue among drivers and longer waiting times for travellers.

 Our findings suggest that profit maximization is an effective strategy for ride-sharing platforms to increase their overall revenue and efficiency. However, platforms should be aware of the potential trade-offs associated with this strategy.  The findings of this study can inform policy makers who are developing regulations for ride-sharing platforms. For example, policy makers may want to consider requiring ride-sharing platforms to implement minimum wage requirements for drivers or to limit the amount of time that travellers can be kept waiting. In general, the findings of this study suggest that ride-sharing platforms should carefully consider the trade-offs associated with different pricing strategies when choosing how to set their prices. Platforms should also be aware of the potential impact of their pricing strategies on drivers, travellers, and society as a whole.

This study provides a valuable starting point for understanding the impact of different pricing strategies on ride-sharing platforms. However, there are several areas where future research could be conducted to build on the findings of this study. First, future research could investigate the impact of different pricing strategies on other important outcomes, such as driver satisfaction, travellers satisfaction, and traffic congestion. Second, future research could explore the impact of different pricing strategies in different markets and under different conditions. Third, future research could develop and test new pricing strategies that are designed to balance the needs of drivers, travellers, and the platform itself.

\section{Conclusion}

Ride-pooling has emerged as a promising solution to urban transportation challenges, aiming to improve efficiency, reduce traffic congestion, and provide cost-effective travel options. In this paper, we have investigated the impact of different pricing strategies on ride-pooling platforms, considering the dynamics between drivers, travellers, and the platform. By employing an agent-based framework (MaaSSim) and simulating various scenarios in the city of Delft, The Netherlands, we evaluated three distinct pricing policies: profit maximization, pooled rides, and solo rides.

The experiments revealed significant insights into the performance of each pricing strategy. The profit maximization strategy proved to be highly efficient in terms of service rates and revenue generation. It maximized the engagement of drivers and increased the platform's overall revenue at the cost of slightly longer waiting times for travellers. However, this approach exhibited an uneven distribution of revenue among drivers.

Conversely, the pooled ride strategy, while generating less overall revenue, ensured a more equitable distribution of earnings among drivers. Additionally, it offered a sustainable source of commission for the platform, promoting stability. Travellers experienced longer waiting times compared to solo rides, but the strategy prioritized travellers comfort and ride-sharing experience. As compared with the pooled rides, the traditional solo rides offered the shortest waiting time for the travellers. 

%Solo rides, representing a traditional ride-hailing approach offered the shortest waiting times for travellers.

The occupancy analysis emphasized the potential for efficient resource utilization, with profit maximization achieving higher occupancy levels. On the other hand, the pooled ride strategy, while prioritizing passenger comfort and reducing wait times, had lower occupancy rates.

In conclusion, selecting an appropriate pricing strategy in ride-pooling platforms is a delicate balance between maximizing revenue, ensuring traveller satisfaction, and engaging drivers effectively. The profit maximization strategy highlighted the revenue potential and resource optimization, though it required careful consideration of waiting times. Pooled rides, while slightly less lucrative, provided a more stable and equitable revenue distribution, aligning with long-term platform sustainability. %Solo rides, though less profitable, offered an 

%%
%% The next two lines define the bibliography style to be used, and
%% the bibliography file.
\bibliographystyle{ACM-Reference-Format}
\bibliography{sample-base}

%%
%% If your work has an appendix, this is the place to put it.

\end{document}